%% file: zprime_body.tex
\documentclass[aps,prl,twocolumn,showpacs,superscriptaddress,groupedaddress]{revtex4}  
\pdfoutput=1
\usepackage[utf8]{inputenc}
\usepackage[T1]{fontenc}
\usepackage{lineno}
\usepackage{graphicx} 
\usepackage{dcolumn}  
\usepackage{colordvi}
\usepackage{color}
\usepackage{amssymb}
\usepackage{amsmath}
\usepackage{url}
\graphicspath{{ps}}
\usepackage{hyperref}
\usepackage{tabularx}
\usepackage{xcolor}
\usepackage{gensymb}

\renewcommand{\figurename}{Fig.}
\renewcommand{\tablename}{Table}

\input ./zprime_symbols.tex

\newcommand {\lmultau} {$L_{\mu}-L_{\tau}$}

\newcommand {\zprime} {$Z^\prime$}

\begin{document}

\title {Search for an invisibly decaying $Z^{\prime}$ boson at Belle~II in $e^+ e^- \to \mu^+ \mu^- (e^{\pm} \mu^{\mp})$ plus missing energy final states}

\input{./zprime_authors.tex}


\begin{abstract}

Theories beyond the standard model often predict the existence of an additional neutral boson, the \zprime. 
Using data collected by the Belle~II experiment during 2018 at the SuperKEKB collider, we perform the first searches for the invisible decay of a \zprime\ in the process $e^+ e^- \to \mu^+ \mu^- Z^{\prime}$ and of a lepton-flavor-violating \zprime\ in $e^+ e^- \to e^{\pm} \mu^{\mp} Z^{\prime}$.
We do not find any excess of events and set 90$\%$ credibility level upper limits on the cross sections of these processes.
We translate the former, in the framework of an \lmultau\ theory, into upper limits on the \zprime\ coupling constant at the level of $5 \times 10^{-2}$ -- 1 for $M_{Z^\prime}\leq 6 \gevcc$.

\end{abstract}

\maketitle


The standard model (SM) is a successful and highly predictive theory of fundamental particles and interactions. 
However, it cannot be considered a complete description of nature, as it does not account for many phenomena, including dark matter. 

The \lmultau\ extension of the SM~\cite{Shuve:2014doa, Altmannshofer:2016jzy} gauges the difference of the leptonic muon and tau number,  giving rise to a new vector boson, the \zprime. 
The \zprime\ couples to the SM only through the $\mu$, $\tau$, $\nu_\mu$ and $\nu_\tau$, with coupling constant $g^{\prime}$.
The \lmultau\ model is potentially able to address important open issues in particle physics, including the anomalies in the $b\rightarrow s \mu^+\mu^-$ decays reported by the LHCb experiment~\cite{lhcb}, the anomaly in the muon anomalous magnetic moment $(g-2)_{\mu}$~\cite{g-2}, and dark matter phenomenology, if extra matter is charged under \lmultau~\cite{Shuve:2014doa, Curtin:2014cca}. 
We investigate here, for the first time, the specific invisible decay topology $e^+e^- \to \mu^+\mu^- Z^{\prime}$, $Z^{\prime} \to \text{invisible}$, where the \zprime\ production occurs via radiation off a final state muon.
The decay branching fractions (BF) to neutrinos are predicted to vary between 33\% and 100\% depending on the $Z^{\prime}$ mass~\cite{Curtin:2014cca}.
This model (``standard \zprime'' in the following) is poorly constrained at low masses.
Related searches have been performed by the \babar\ and CMS experiments for a $Z^{\prime}$ decaying to muons~\cite{TheBABAR:2016rlg, cms}.
Our search is, therefore, the first to have some sensitivity to \zprime\ masses $m_{Z^\prime} < 2 m_\mu$. 
If the \zprime\ is able to decay directly into a pair of dark matter particles $\chi \bar{\chi}$, one assumes BF$(Z^{\prime} \to \chi \bar{\chi}) \approx 1$ due to the expected much stronger coupling relative to SM particles.
We provide separate results for this scenario, which is not constrained by existing measurements. 

The second scenario we consider postulates the existence of a lepton-flavor-violating (LFV) boson, either a scalar or a vector (``LFV \zprime'' in the following), which couples to leptons~\cite{Galon2017, Galon2017:2}.
We focus on the LFV $e-\mu$ coupling.
While the presence of LFV mediators can be constrained by  measurements of the forward-backward asymmetry in  $e^+e^- \to \mu^+\mu^-$~\cite{Galon2017:2, TheDELPHICollaboration2006}, we present here a direct, model-independent search of $e^+e^- \to e^{\pm}\mu^{\mp} Z^{\prime}$, $Z^{\prime} \to \text{invisible}$.
The presence of missing energy decays make these searches especially suitable for an \epem collider. 

The Belle~II detector~\cite{Abe:2010sj} operates at the SuperKEKB electron-positron collider~\cite{superkekb}, located at the KEK laboratory in Tsukuba, Japan. 
Data were collected at the center-of-mass (CM) energy of the $\Upsilon$(4S) resonance from April to July 2018.
The energies of the electron and positron beams are $7\gev$ and $4\gev$, respectively, resulting in a boost of $\beta\gamma = 0.28$ of the CM frame relative to the lab frame. 
The integrated luminosity used in this analysis amounts to 276\invpb~\cite{lumi}.

The Belle~II detector consists of several subdetectors arranged around the beam pipe in a cylindrical structure. A superconducting solenoid, situated outside of the calorimeter, provides a 1.5~T magnetic field. Subdetectors relevant for this analysis are briefly described here; a description of the full detector is given in~\cite{Abe:2010sj, ref:b2tip}. 
The innermost subdetector is the vertex detector (VXD), which includes two layers of silicon pixels and four outer layers of silicon strips.
Only a single octant of the VXD was installed during the 2018 operations~\cite{paladino}. 
The main tracking device (CDC) is a large helium-based small-cell drift chamber.
The electromagnetic calorimeter (ECL) consists of a barrel and two endcaps made of CsI(Tl) crystals.
The $z$ axis of the laboratory frame is along the detector solenoidal axis in the direction of the electron beam.
``Longitudinal'' and ``transverse'' are defined with respect to this direction, unless otherwise specified.

The invisible \zprime\ signature is a peak in the distribution of the invariant mass of the system recoiling against a lepton pair. ``Recoil'' quantities such as mass and momentum refer to this system. 
They coincide with \zprime\ properties in the case of signal events, and typically correspond to undetected SM particles in the case of background events.  
The analysis uses events with exactly two tracks, identified as $\mu\mu$ or $e \mu$, and minimal other activity in the ECL. 
The standard \zprime\ selection is optimized using simulated events prior to examining data; the same criteria, aside from an electron in the final state, are used for the LFV \zprime\ search.
The dominant backgrounds are SM final states with missing energy and two tracks identified as leptons.
These are radiative muon pairs ($e^+e^- \rightarrow \mu^+\mu^-\gamma (\gamma)$) with one or more photons which are not detected due to inefficiency or acceptance, $e^+e^- \rightarrow \tau^+\tau^-(\gamma)$, and $e^+e^- \rightarrow e^+e^-\mu^+\mu^-$ with electrons outside the acceptance.
Control samples are used to check background rates predicted by simulation and to infer correction factors and related uncertainties. 
Upper limits on the standard \zprime\ cross section are computed with a counting technique in windows of the recoil mass distribution.
For the LFV \zprime\ model-independent search, upper limits are interpreted in terms of signal efficiency times cross section. 
Details of each of these steps are described below. 

Signal events are generated with \texttt{MadGraph 5}~\cite{Alwall2014} for standard $Z^{\prime}$ masses ranging from 0.5 to $8\gevcc$ in steps of $0.5\gevcc$.
The following background sources are generated using the specified generators: 
$e^+e^- \rightarrow \mu^+\mu^-(\gamma)$ (\texttt{KKMC}~\cite{ref:kkmc});
$e^+e^- \rightarrow \pi^+\pi^-(\gamma)$ (\texttt{PHOKHARA}~\cite{ref:phokhara});
$e^+e^- \rightarrow e^+e^-(\gamma)$ (\texttt{BabaYaga@NLO}~\cite{ref:babayaga});
$e^+e^- \rightarrow \tau^+\tau^-(\gamma)$ (\texttt{KKMC}~\cite{ref:kkmc} with \texttt{TAUOLA}~\cite{ref:tauola});
$e^+e^- \rightarrow e^+e^-\mu^+\mu^-$; and   
$e^+e^- \rightarrow e^+e^-e^+e^-$ (\texttt{AAFH}~\cite{ref:fourlepton}). 
The detector geometry and the interactions of the final state particles with the material are simulated using \texttt{\textsc{Geant4}}~\cite{ref:geant4} and the Belle~II Analysis Software Framework~\cite{basf2}.

The standard \zprime\ search uses the CDC two-track trigger, which selects events with at least two tracks with an azimuthal opening angle larger than $90^\circ$.
The LFV \zprime\ search uses the ECL trigger, which selects events with total energy in the barrel and part of the endcap above $1\gev$.
Both triggers reject events that are consistent with being Bhabha scatterings. 

To reject spurious tracks and beam induced background, ``good'' tracks are required to have transverse and longitudinal projections of the distance of closest approach with respect to the interaction point smaller than 0.5\cm and 2.0\cm, respectively.
Photons are classified as ECL clusters with energy greater than $100\mev$, which are not associated with tracks. 
Quantities are defined in the laboratory frame unless specified otherwise. Events are required to pass the following selection criteria.
\begin{enumerate}
\item  Exactly two oppositely charged good tracks, with polar angles in a restricted barrel ECL acceptance $\theta\in[37,120]^\circ$ and with azimuthal opening angle $>90^\circ$, to match the CDC trigger requirement.
\item Recoil momentum pointing into the ECL barrel acceptance $\theta\in[32,125]^\circ$, to exclude inefficient regions where photons from radiative backgrounds can escape undetected.
This selection is applied only for recoil masses below $3\gevcc$; missed radiative photons are unlikely to produce higher masses.
\item An ECL-based particle identification (PID) selection: $0.15<E<0.4\gev$ and $E/pc<0.4$ for muons; $0.8<E/pc<1.2$ and $E>1.5\gev$ for electrons, where $E$ is the energy of the ECL cluster associated to a track of momentum $p$.  
\item No photons within a $15^\circ$ cone around the recoil momentum direction in the CM frame, to suppress radiative lepton pair backgrounds.
\item Total photon energy less than $0.4\gev$ and no $\pi^0$ candidates (pairs of photons with invariant masses within $10\mevcc$ of the nominal \piz value)
\end{enumerate}
After this selection, the background for recoil masses below $7\gevcc$ is dominated by $e^+e^- \rightarrow \tau^+\tau^-(\gamma)$ events with $\tau\rightarrow \mu$, or $\tau\rightarrow \pi$ where the pion is misidentified as a muon.
 
In subsequent steps of the analysis, events are grouped into windows of recoil mass.
The width of these windows is $\pm 2 \sigma$, where $\sigma$ is the recoil mass resolution.
It is obtained by fitting each \zprime\ recoil mass distribution with a sum of a Crystal Ball (CB)~\cite{cb1, cb2, cb3} and a Gaussian function with coincident peaks.
The resolution is computed as the sum in quadrature of the CB and Gaussian widths weighted according to their contributions.
The choice of $\pm 2\sigma$ maximizes a figure of merit (FOM)~\cite{Punzi:2003bu} over the full spectrum.
Mass window widths vary from $1150\mevcc$ at $M_{Z^{\prime}} = 0.5 \gevcc$ to a minimum of $51\mevcc$ at $M_{Z^{\prime}} = 6.9 \gevcc$.
There are in total 69 mass windows below $8\gevcc$. 

Studies with radiative muon pair events ($\mu\mu\gamma$ sample) indicate that the recoil mass widths for data and simulation are consistent. No systematic uncertainty is assigned.

\begin{figure}[!htb]
  \centering
  \includegraphics[width=0.99\columnwidth]{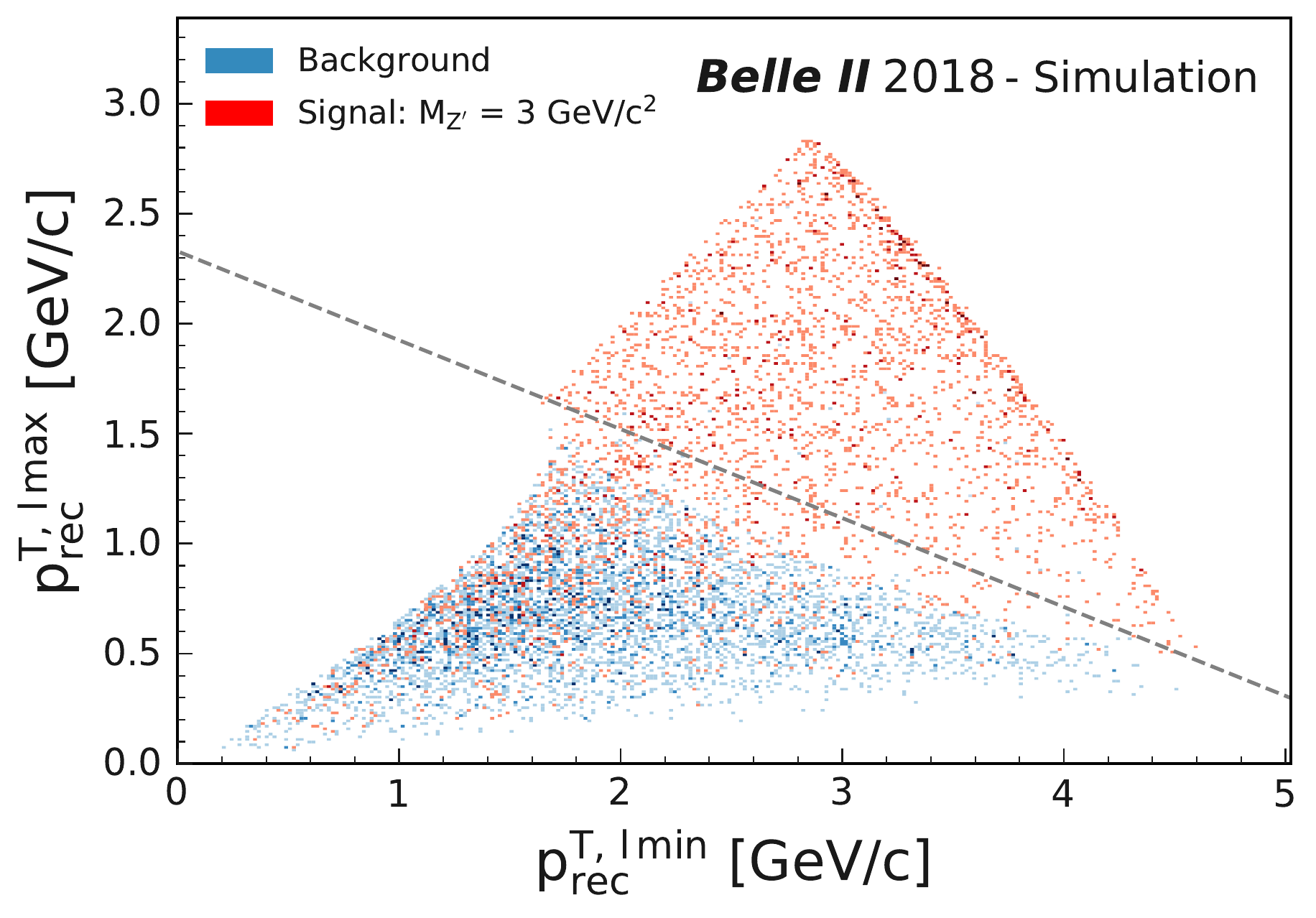}
  \caption{$p^{\rm T,lmax}_{\rm rec}$ vs.\ $p^{\rm T,lmin}_{\rm rec}$ distributions after the optimal $p^{\rm T}_{\mu\mu}$ selection for $M_{Z^{\prime}} = 3\gevcc$ signal (red) and for background (blue).
  $p^{\rm T,lmax}_{\rm rec}$ ($p^{\rm T,lmin}_{\rm rec}$) is the transverse recoil momentum with respect to the direction of the muon with maxiumum (minimum) momentum in the CM frame. 
  The optimal separation line is superimposed.}
  \label{fig:LP-tau}
\end{figure}

A final selection, denoted as ``$\tau$ suppression'', exploits the kinematics of the \zprime\ production,  which occurs radiatively from a final state muon, to further suppress \tautau events in which the missing momentum arises from neutrinos from both $\tau$ decays. 
The variables, defined in the CM frame, are: the transverse recoil momentum with respect to the lepton with the higher momentum $p^{\rm T,lmax}_{\rm rec}$; with respect to the lower momentum  $p^{\rm T,lmin}_{\rm rec}$; the transverse momentum of the dilepton pair ($p^{\rm T}_{\mu\mu}$ or $p^{\rm T}_{e\mu}$).
Figure~\ref{fig:LP-tau} shows $p^{\rm T,lmax}_{\rm rec}$ versus  $p^{\rm T,lmin}_{\rm rec}$ for a standard \zprime\ mass of $3\gevcc$ and for the total simulated background in the corresponding recoil mass window.

For the standard \zprime\ search, a linear cut is imposed in the $p^{\rm T,lmax}_{\rm rec}$--$p^{\rm T,lmin}_{\rm rec}$ plane and a simultaneous selection $p^{\rm T}_{\mu\mu} > p^{\rm T}_{\rm cut}$ where the cut values are determined using an optimization procedure that numerically maximizes the FOM in each recoil mass window.
$p^{\rm T}_{\rm cut}$ is typically 1.5--2.0\gevc and is effective in suppressing the remaining $\mu^+\mu^-(\gamma)$ and $\epem\mumu$ backgrounds.  
For masses higher than $7\gevcc$, signal and background overlap in the $p^{\rm T,lmax}_{\rm rec}$--$p^{\rm T,lmin}_{\rm rec}$ plane and effective separation lines are not found. 
The same values are used for the LFV \zprime\ search. 

Trigger, tracking and particle identification efficiencies are studied on control samples.
The performance of the CDC two-track trigger is studied on data samples, mostly radiative Bhabha scattering  events, selected by means of the ECL trigger.
The efficiency is $(79 \pm 5)\%$ when both tracks are within the acceptance of selection 1; the uncertainty is systematic and is due to kinematic dependencies.
The performance of the ECL trigger is studied using $e^+e^-\to\mu^+\mu^- \gamma$ events with $E_\gamma>1\gev$ that are selected with the CDC two-track trigger.
The efficiency is found to be uniformly $(96\pm1)\%$ in the ECL barrel region.

The tracking efficiency for data is compared to simulation using radiative Bhabha and  $e^+e^-\to\tau^+\tau^-$ events.
Differences are found to be $10\%$ for two-track final states.
A 0.90 correction factor is applied to simulation, with a $4\%$ systematic uncertainty due to kinematic dependencies.

The PID efficiency for data is compared to simulation using samples of four-lepton events from two-photon mediated processes.
Discrepancies at the level of $2\%$ per track are found, resulting in a systematic uncertainty of $4\%$. 

The selection criteria before the $\tau$ suppression are studied using signal-free control samples in data and simulation.
We use the $\mu\mu\gamma$ sample defined above and an analogously defined $e\mu\gamma$ sample to check the low recoil mass region. Kinematic quantities are computed without taking into account the presence of the photon. 
We also select $\mu\mu$ and $e\mu$ samples that satisfy requirements 1--5, but which fail the $p^{\rm T,lmax}_{\rm rec}$--$p^{\rm T,lmin}_{\rm rec}$ requirement.
These studies indicate that, factoring out the 0.90 tracking efficiency correction, the
efficiency before the $\tau$ suppression is 25\% lower for $\mu^+ \mu^-$ events in data than in simulation, but agrees for $e^{\pm} \mu^{\mp}$ events.  
A variety of studies failed to uncover the source of this discrepancy, which is consistently found to be independent of all checked quantities, including the recoil mass. 
The background predictions from simulation and the signal efficiency are thus corrected with a scaling factor of 0.75 for $\mu^+ \mu^-$ events.
After the inclusion of these corrections, the background level before the $\tau$ suppression selection agrees with the simulation in both samples within a 2\% statistical uncertainty~\cite{supplemental}, which is used as a systematic contribution.
This is a strong constraint for the standard \zprime\ signal efficiency as well, as the topology of background and signal events (a pair of muons and missing energy) is identical for signal and background and the discrepancy in the measured yield is found not to depend on kinematic quantities (see above).
Nevertheless, we conservatively assign a systematic uncertainty of $12.5\%$ on the correction factor to the signal efficiency for the dimuon sample, half the size of the observed discrepancy.

To study the $\tau$ suppression, we use an \epem sample selected using the same analysis criteria, but with both tracks satisfying the electron criteria in selection 3.
The resulting sample includes $e^+e^-\gamma$, $e^+e^-e^+e^-$ and \tautau events where both leptons decay to electrons. The latter has the same kinematic features of the most relevant background source to both searches.
Agreement between data and simulation is found after the $\tau$ suppression, within a 22\% statistical uncertainty. This is taken as a systematic uncertainty on the background; no systematic uncertainty due to this effect is considered for the signal, as the selection has a high efficiency (around 50\%, slightly depending on the \zprime\ mass), and the distributions on which it is based are well reproduced in simulation. 

After the corrections for the two-track trigger efficiency and for the data/simulation discrepancy in \mumu events, signal efficiencies are found to range between 2.6\% and 4.9\% for \zprime\ masses below $7\gevcc$. 
Signal efficiencies are interpolated from the generated \zprime\ masses to the center of each recoil mass window.
An additional binning scheme is introduced with a shift of a half bin, to cover hypothetical signals located at the border of two contiguous bins, where the signal efficiency is reduced.  
Systematic uncertainties are summarized in \tablename~\ref{tab:syst}.

\begin{table}[!htb]
  \caption{Relative systematic uncertainties affecting the $\mu^+ \mu^-$ and $e^{\pm} \mu^{\mp}$ analyses.} 
  \centering
  \begin{tabular}{ccc}
    \hline \hline 
    Source & $\mu^+ \mu^-$ & $e^{\pm} \mu^{\mp}$\\
    \hline
    Trigger efficiency & 6\% & 1\% \\
    Tracking efficiency & 4\% & 4\% \\
    PID & 4\% & 4\% \\
    Luminosity & 0.7\% & 0.7\% \\      
    $\tau$ suppression (background)  & 22\% & 22\% \\
    Background before $\tau$ suppression & 2\% & 2\% \\
    Discrepancy in $\mu\mu$  yield (signal) & 12.5\% & -- \\      
    \hline \hline
  \end{tabular}
  \label{tab:syst}
\end{table}

\begin{figure}[!htb]
  \centering
  \includegraphics[width=0.99\columnwidth]{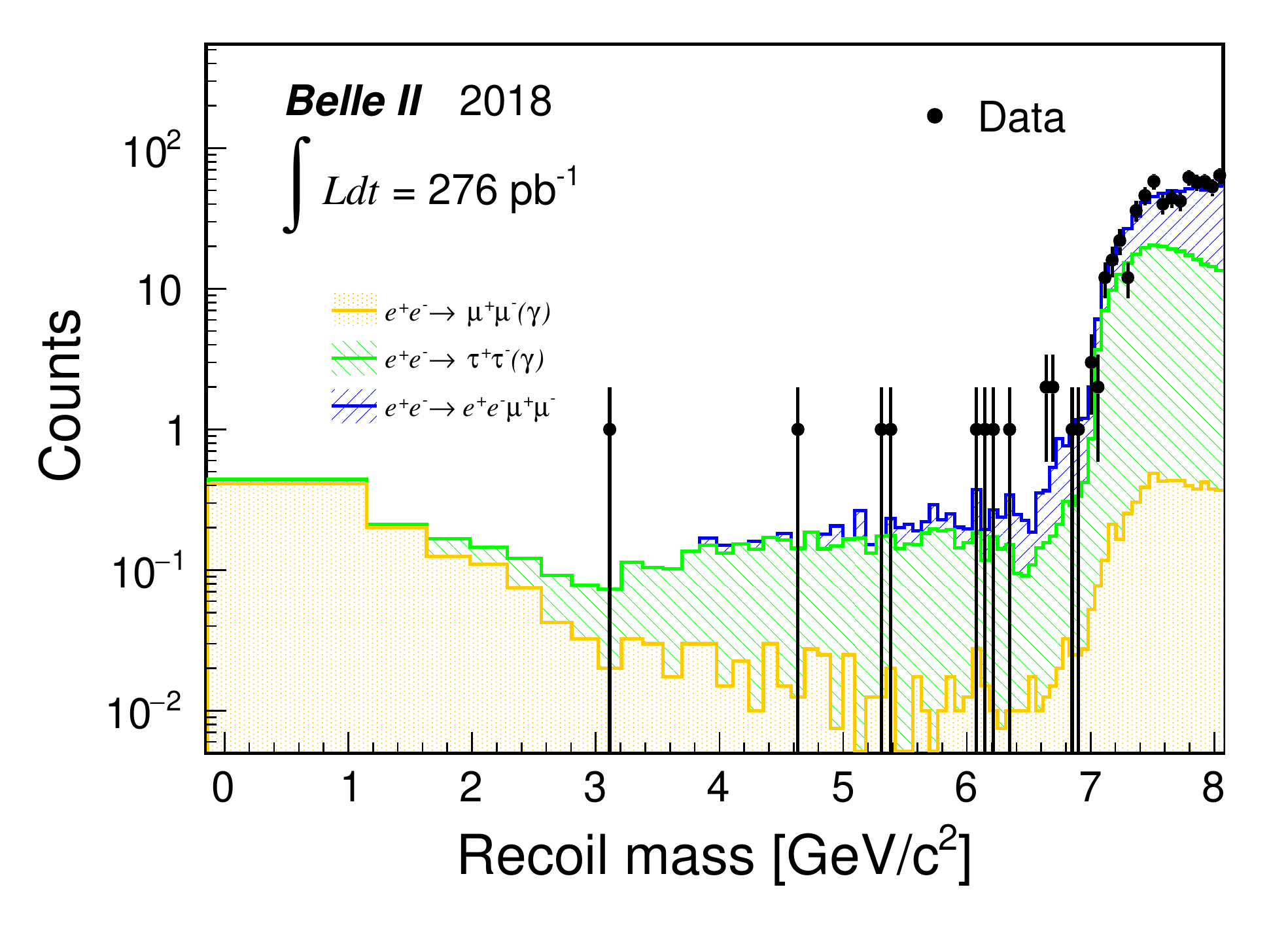}
  \caption{Recoil mass spectrum of the \mumu sample.
   Simulated samples (histograms) are rescaled for luminosity, trigger (0.79), and tracking (0.90) efficiencies, and the correction factor (0.75, see text).
   Histogram bin widths indicate the recoil mass windows.}
  \label{fig:LP-mumu_after}
\end{figure}

\begin{figure}[!htb]
  \centering
  \includegraphics[width=0.99\columnwidth]{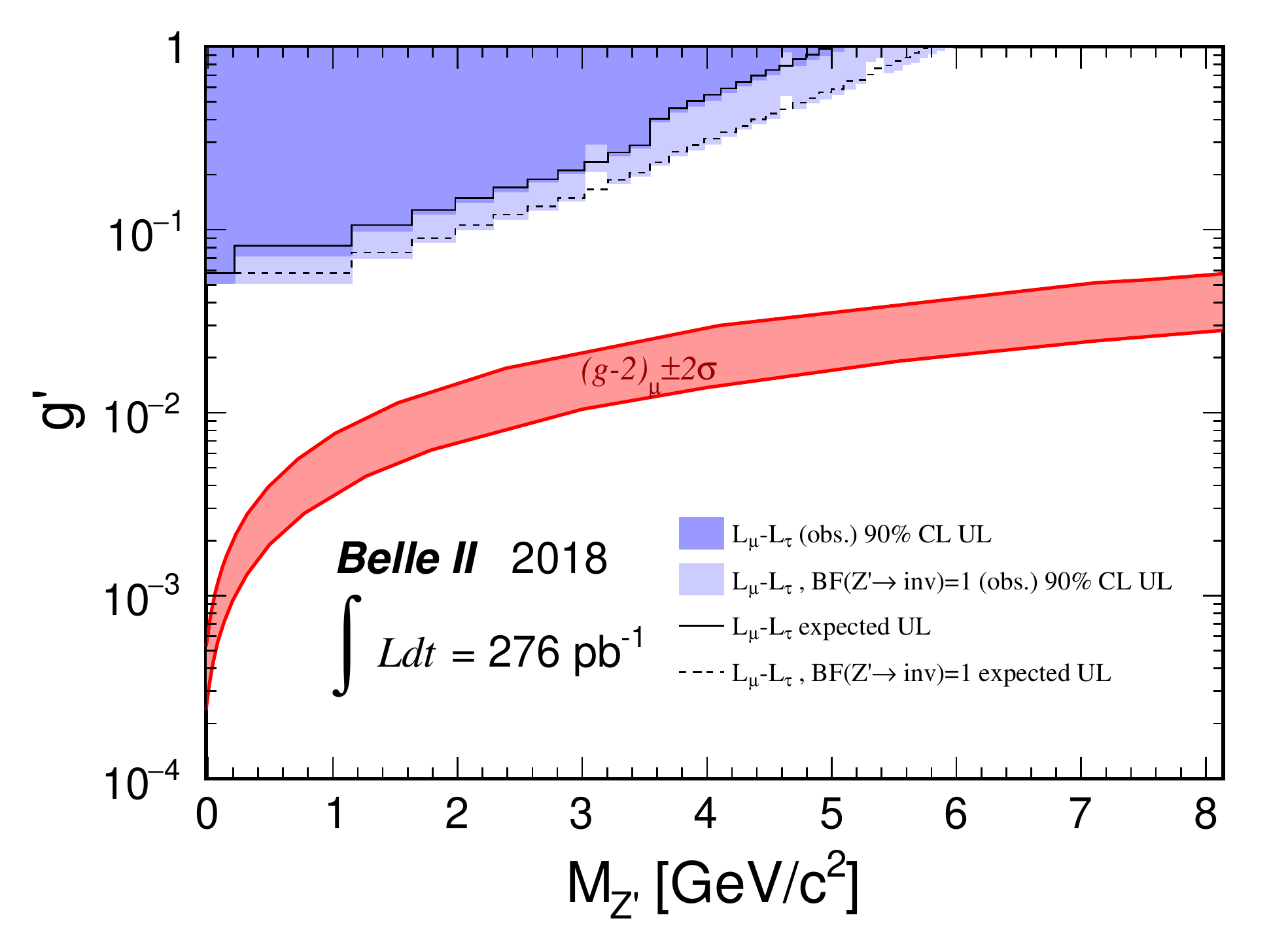}
  \caption{90$\%$ CL upper limits on coupling constant  $g^{\prime}$.
  Dark blue filled areas show the exclusion regions for $g^{\prime}$ at 90\% CL, assuming the $L_\mu - L_\tau$ predicted BF for $Z^{\prime} \to \text{invisible}$; light blue areas are for BF$(Z^{\prime} \to \text{invisible})=1$.
  The solid and dashed lines are the expected sensitivities in the two hypotheses. 
  The red band shows the region that could explain the muon anomalous magnetic moment $(g-2)_{\mu} \pm 2\sigma$~\cite{Shuve:2014doa, Curtin:2014cca}. 
  The step at $M_{Z^\prime} = 2m_\mu$ for the $L_\mu - L_\tau$ exclusion region reflects the  change in BF$(Z^{\prime} \to \nu \bar\nu)$.}
  \label{fig:LP-gp_lim-log}
\end{figure}

The final recoil mass spectrum of the \mumu sample is shown in \figurename~\ref{fig:LP-mumu_after}, together with the expected background.
We look for the presence of possible local excesses by calculating for each recoil mass window the probability to obtain a yield greater or equal to that obtained in data given the predicted background, including statistical and systematic uncertainties.
No anomalies are observed, with all results below $3\sigma$ local significance in both the normal and shifted-binning options~\cite{supplemental}.
A Bayesian procedure~\cite{bat_software} is used to compute 90\% credibility level (CL) upper limits on the standard \zprime\ cross section.
We assume flat priors for all positive values of the cross section, while Poissonian likelihoods are assumed for the number of observed and simulated events. Gaussian smearing is used to model the systematic uncertainties. 
Results are cross-checked with log-flat priors and with a frequentist procedure based on the Feldman-Cousins approach~\cite{Feldman:1997qc} and are found to be compatible in both cases~\cite{supplemental}.
Cross section results are translated into 90\% CL upper limits on the coupling constant $g^{\prime}$. 
These are shown in \figurename~\ref{fig:LP-gp_lim-log}, where only values $g^{\prime} \leq 1$ are displayed.
The observed upper limits for models with BF$(Z^{\prime} \to \text{invisible}) < 1$ can be obtained by scaling the light blue curve as $1/\sqrt{\text{BF}}$.

\begin{figure}[!htb]
  \centering
  \includegraphics[width=0.99\columnwidth]{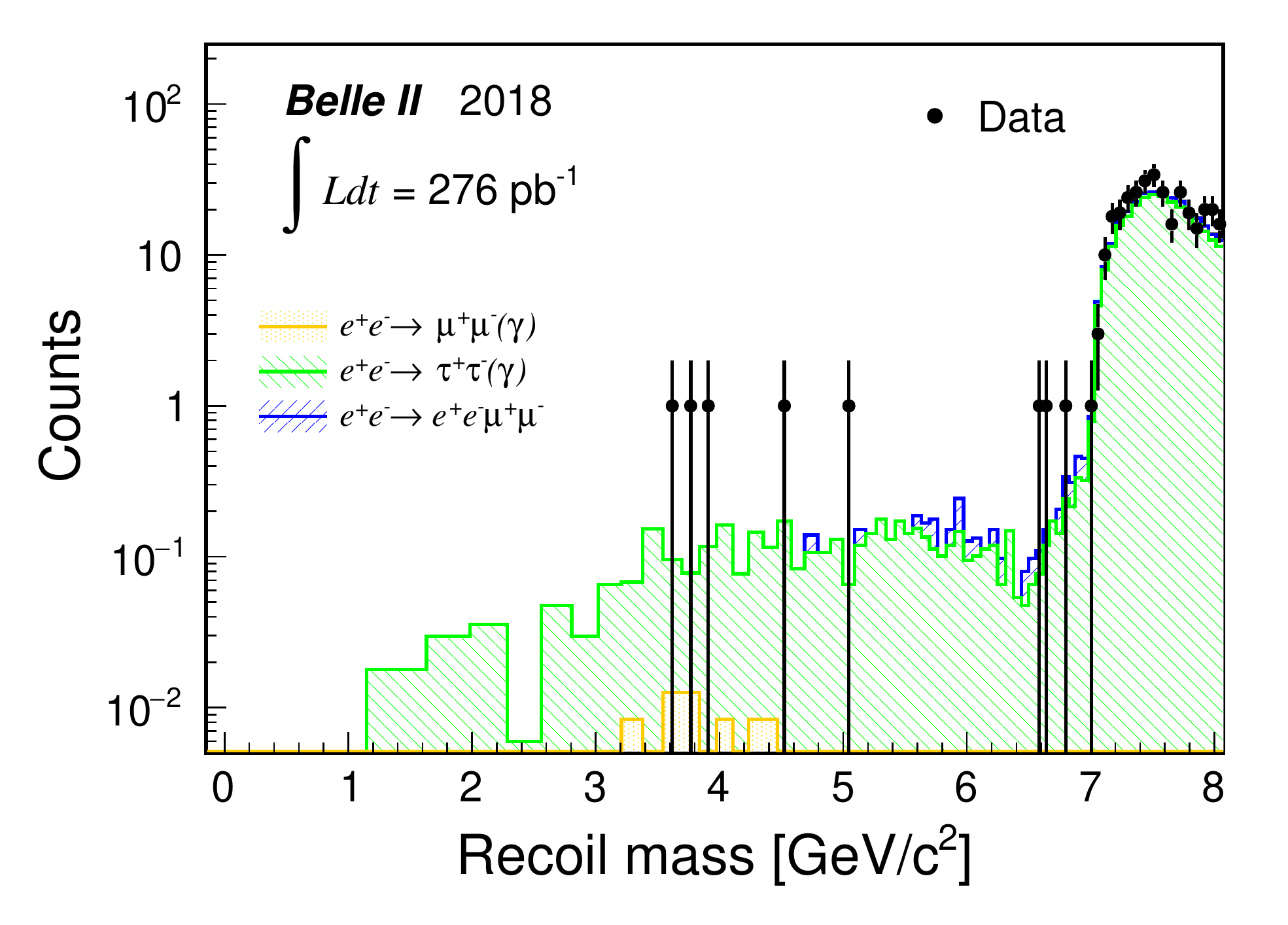}
  \caption{Recoil mass spectrum of the $e^\pm \mu^\mp$ sample.
   Simulated samples (histograms) are rescaled for luminosity, trigger (0.79), and tracking (0.90) efficiencies.
   Histogram bin widths indicate the recoil mass windows.}
  \label{fig:LP-emu_after}
\end{figure}

The final recoil mass spectrum of the $e^\pm \mu^\mp$ sample is shown in \figurename~\ref{fig:LP-emu_after}, together with background simulations.
Again, no anomalies are observed above $3\sigma$ local significance~\cite{supplemental}.
Model-independent 90\% CL upper limits on the LFV \zprime\ efficiency times cross section are computed 
using the Bayesian procedure described above and cross-checked with a frequentist Feldman-Cousins procedure (\figurename~\ref{fig:LP-lim-lfv}). 
Additional plots and numerical results can be found in the supplemental material~\cite{supplemental}. 

\begin{figure}[!htb]
  \centering
  \includegraphics[width=0.99\columnwidth]{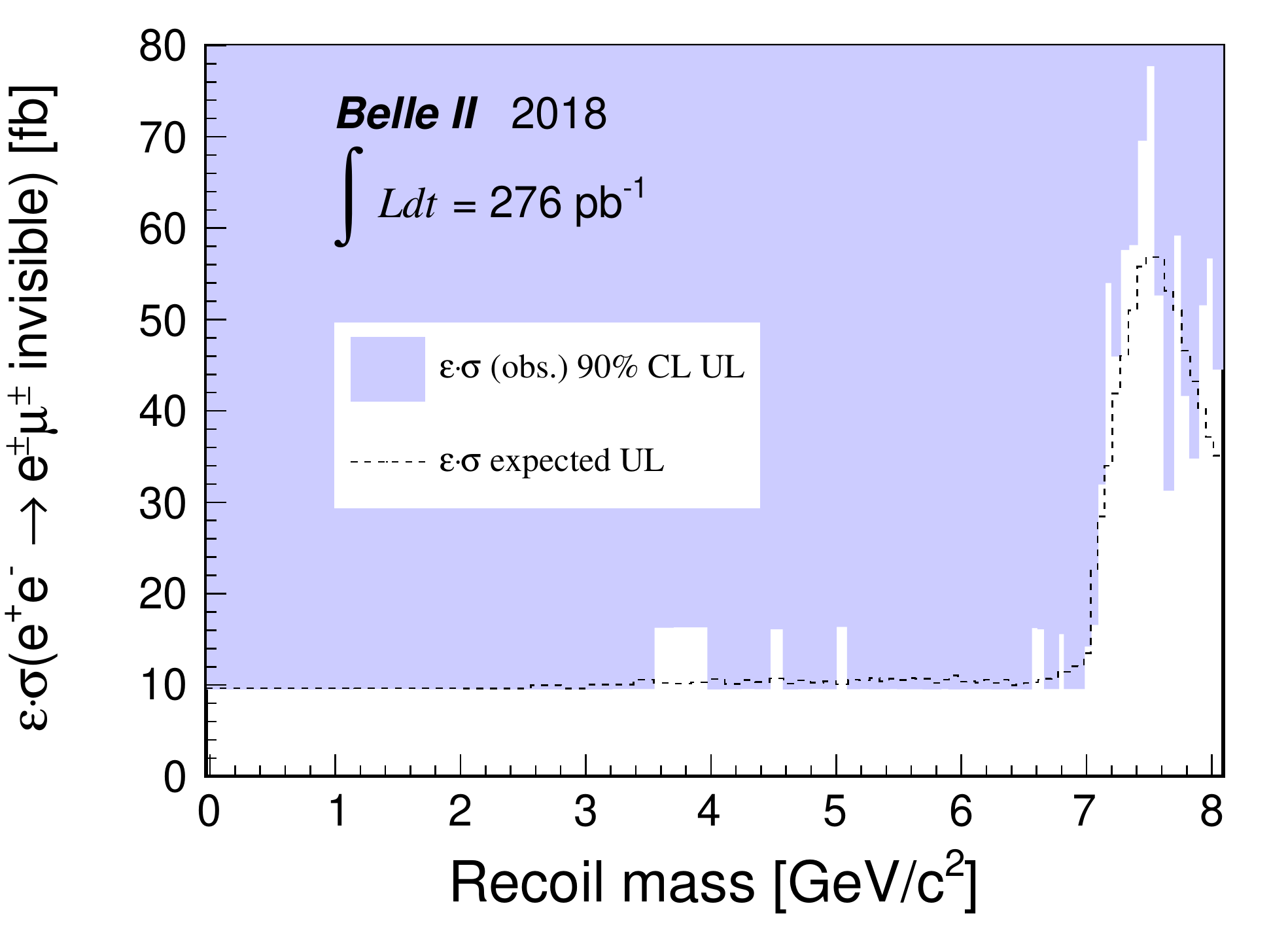}
  \caption{90$\%$ CL upper limits on efficiency times cross section $\epsilon \times \sigma[e^+ e^- \to e^{\pm} \mu^{\mp} \text{invisible}]$. The dashed line is the expected sensitivity.}
  \label{fig:LP-lim-lfv}
\end{figure}

In summary, we have searched for an invisibly decaying \zprime\ boson in the process $e^+e^- \to \mu^+\mu^-Z^{\prime}$ and for a LFV \zprime\ in the process $e^+e^- \to e^{\pm}\mu^{\mp}Z^{\prime}$, using 276\invpb of data collected by Belle~II at SuperKEKB in 2018.
We find no significant excess and set for the first time 90\% CL upper limits on the coupling constant $g^{\prime}$ in the range $5 \times 10^{-2}$ to 1 for the former case and to the efficiency times cross section around 10 fb for the latter. 
The full Belle~II data set, with better muon identification, a deeper knowledge of the detector, and the use of multivariate analysis techniques should be sensitive to the $10^{-3}$ -- $10^{-4}$ $g^{\prime}$ region, where the $(g-2)_\mu$ band currently lies.


\input{./zprime_acknowledgements.tex}

\bibliography{zprime_bibliography}
\bibliographystyle{apsrev4-1}

\end{document}


\title{Search for an invisibly decaying $Z^{\prime}$ boson at Belle II in $e^+ e^- \to \mu^+ \mu^- (e^{\pm} \mu^{\mp})$ plus missing energy final states}

\author{Belle II Collaboration}

\maketitle

This material is submitted as supplementary information for the Electronic Physics Auxiliary Publication Service.

\begin{figure}[!ht]
    \centering
    \includegraphics[width=0.6\linewidth]{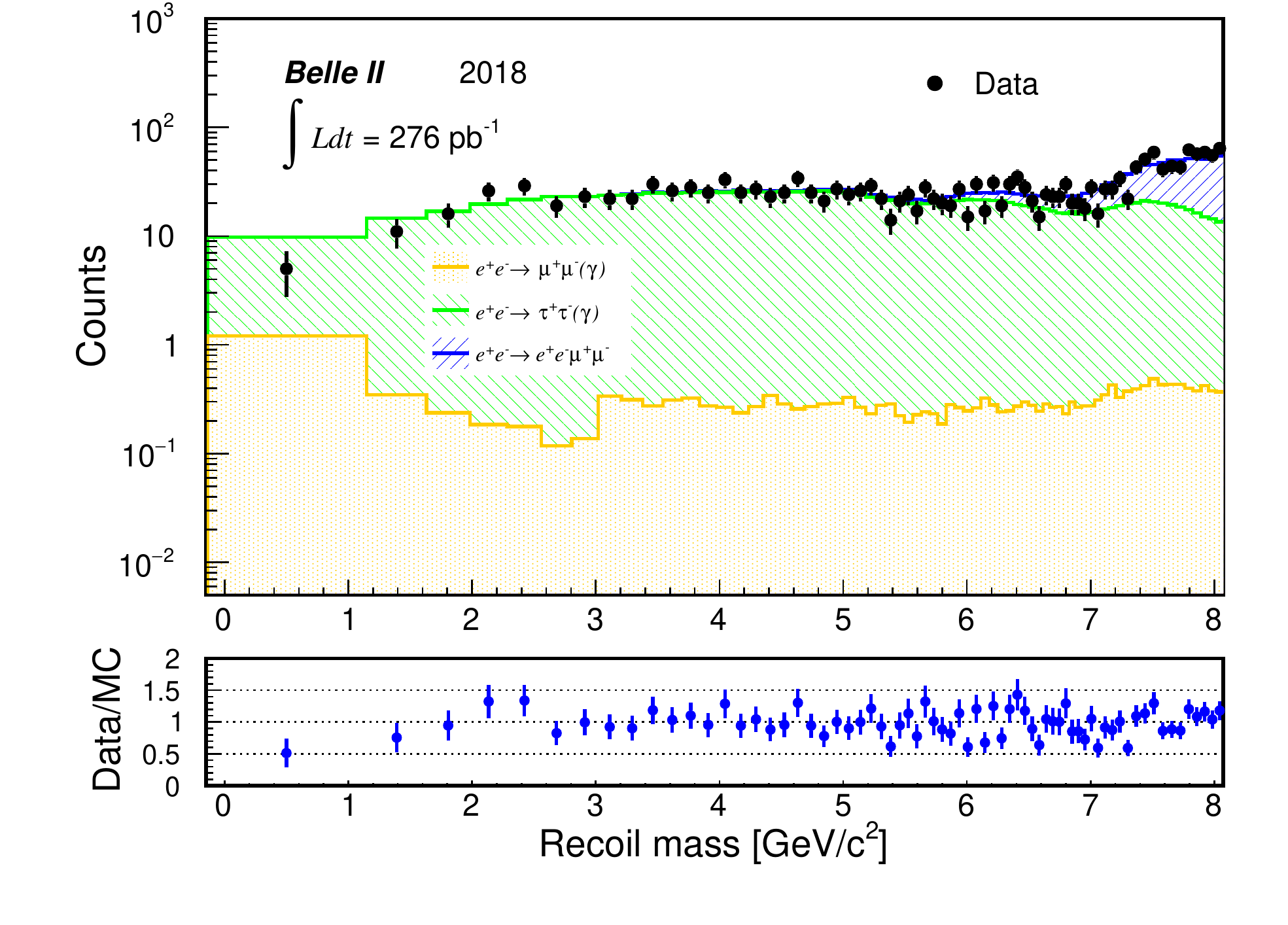}
    \caption{Recoil mass spectrum for the $\mu^+ \mu^-$ sample before the $\tau$ suppression selection.
    Simulated samples (histograms) are rescaled for luminosity, trigger (0.79), and tracking (0.90) efficiencies, and the correction factor (0.75, see text).
    Histogram bin widths indicate the recoil mass windows.}
    \label{fig:fig1}
\end{figure}

\begin{figure}[!ht]
    \centering
    \includegraphics[width=0.6\linewidth]{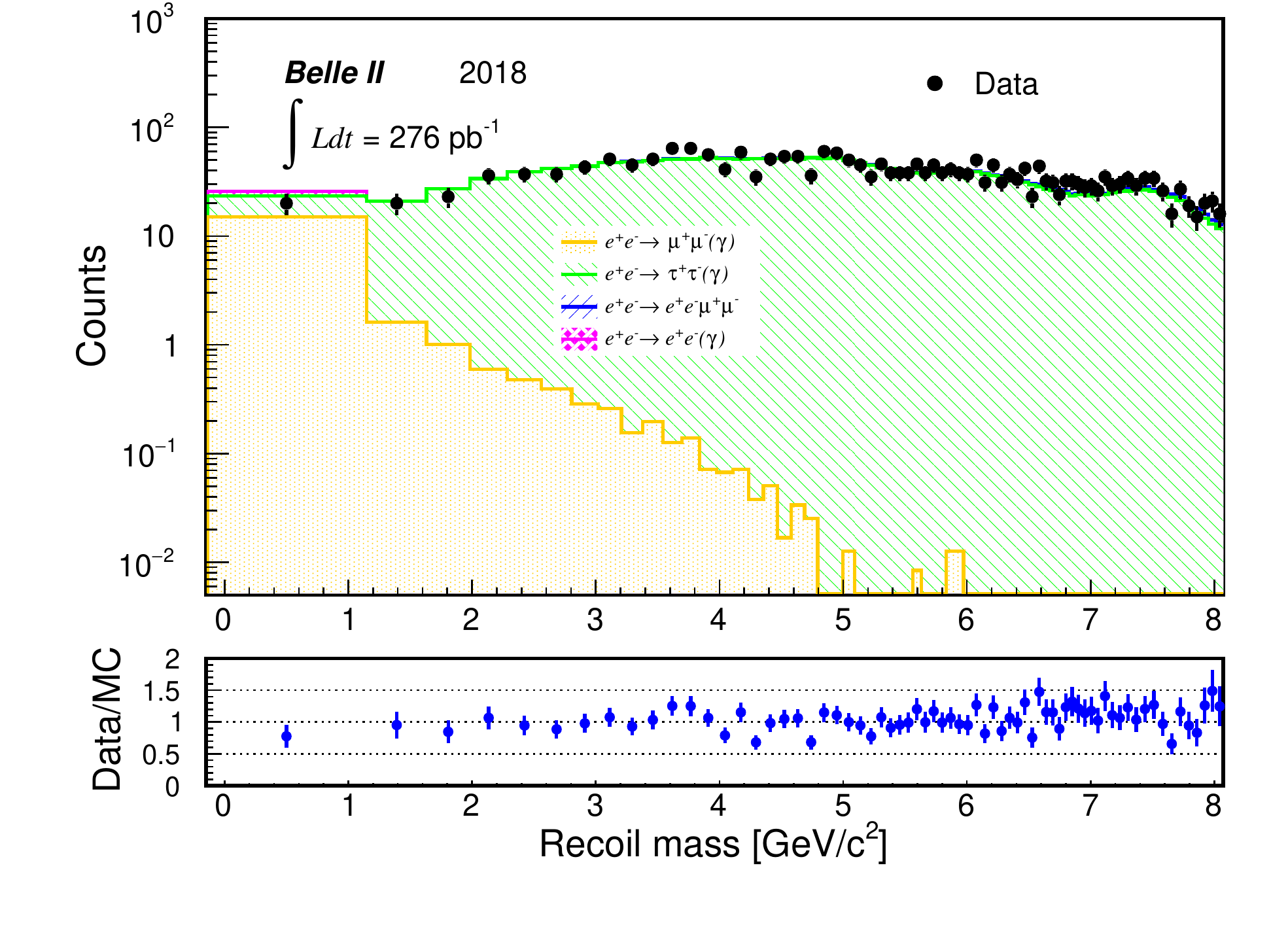}
    \caption{Recoil mass spectrum for the $e^{\pm} \mu^{\mp}$ sample after the $\tau$ suppression selection.
    Simulated samples (histograms) are rescaled for luminosity, trigger (0.79), and tracking (0.90) efficiencies.
    Histogram bin widths indicate the recoil mass windows.}
    \label{fig:fig2}
\end{figure}

\begin{figure}[!ht]
    \centering
    \includegraphics[width=0.6\linewidth]{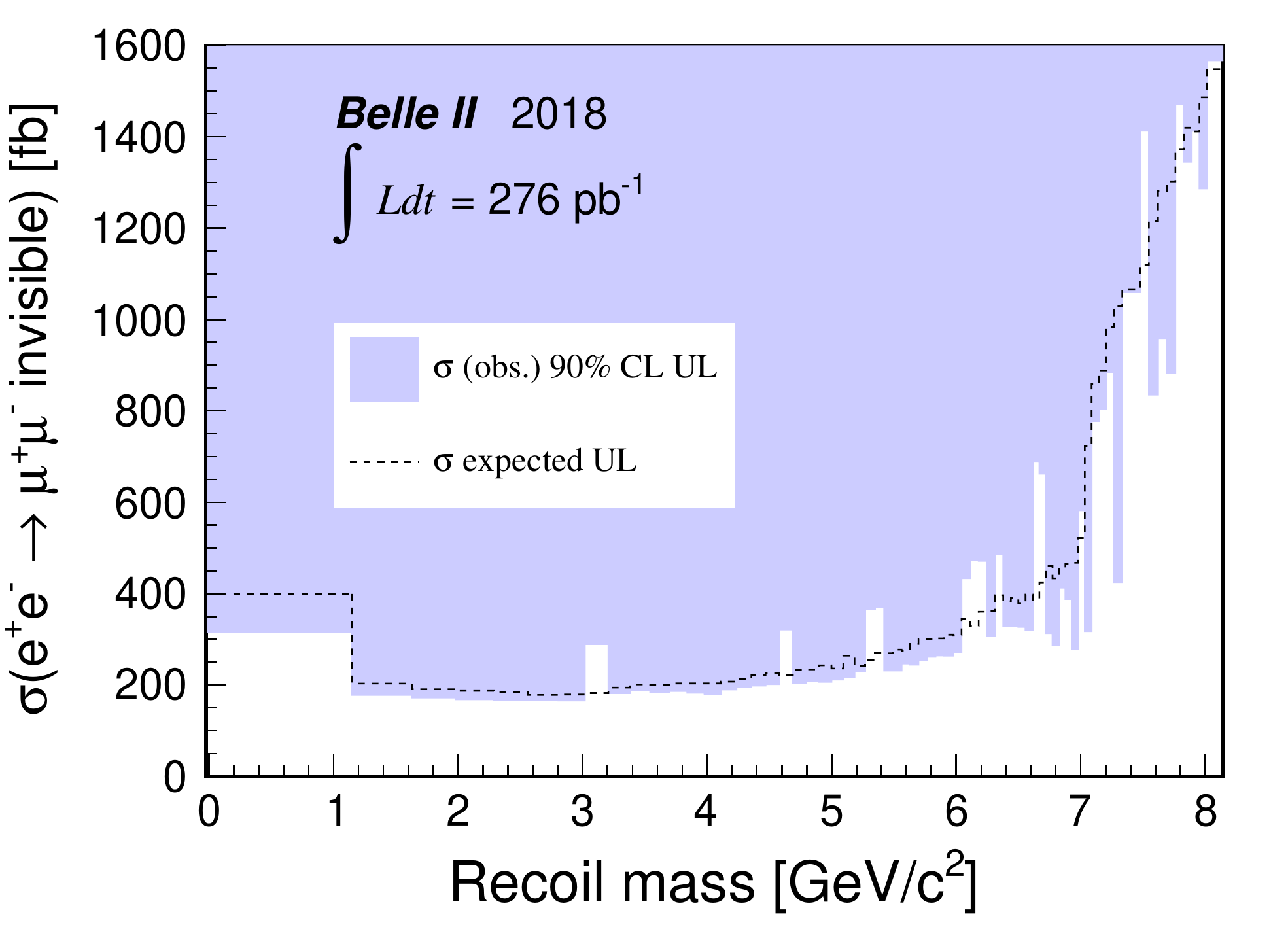}
    \caption{90$\%$ CL upper limits on cross section $\sigma(e^+ e^- \to \mu^+ \mu^- \text{invisible})$.
    The dashed line is the expected sensitivity.}
    \label{fig:fig3}
\end{figure}

\begin{figure}[!ht]
    \centering
    \includegraphics[width=0.6\linewidth]{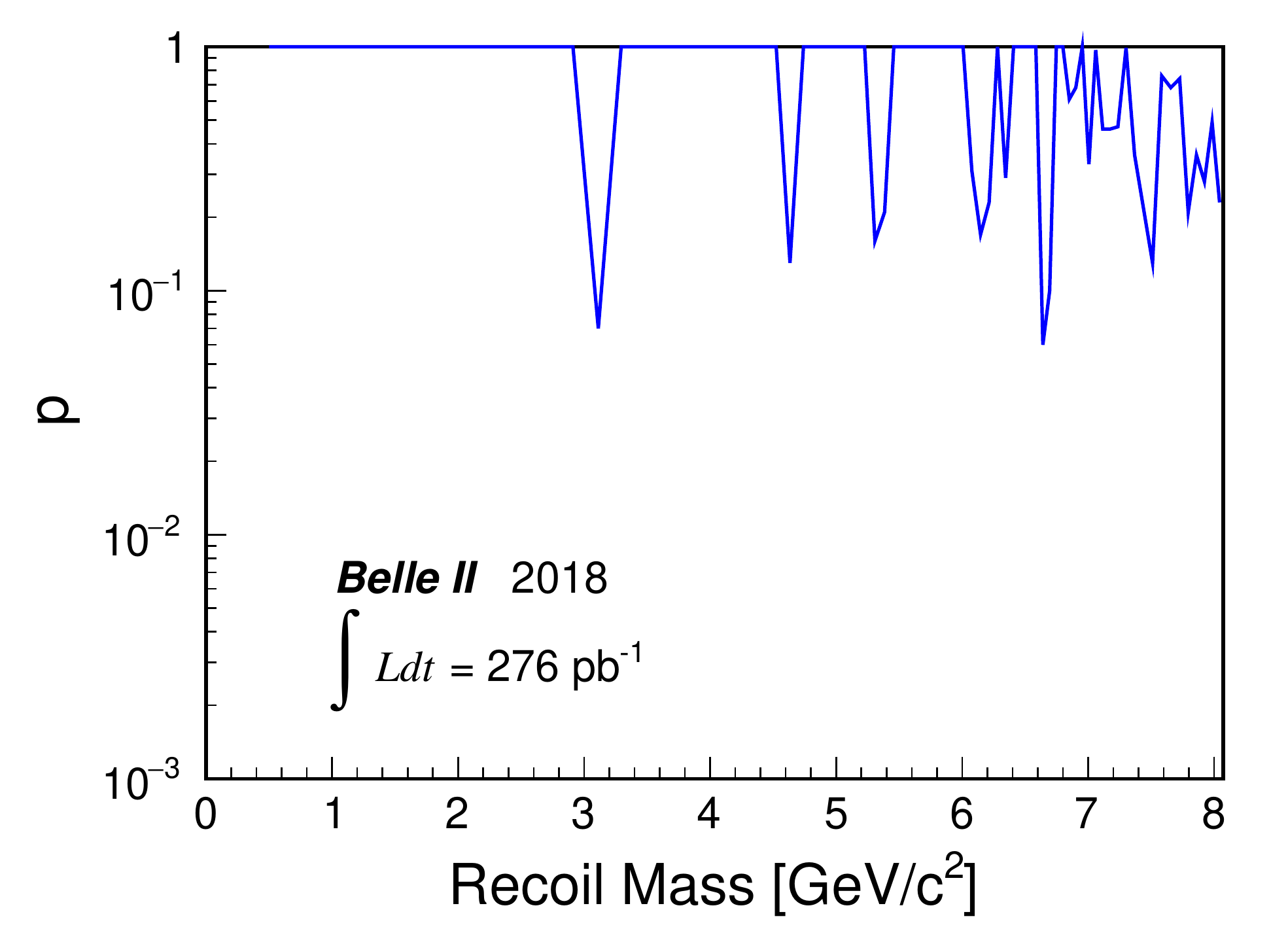}
    \caption{Probability $p$ to observe a number of events greater than the measured one given the predicted background level as a function of the recoil mass for the $\mu^+ \mu^-$ sample.}
    \label{fig:fig4}
\end{figure}

\begin{figure}[!ht]
    \centering
    \includegraphics[width=0.6\linewidth]{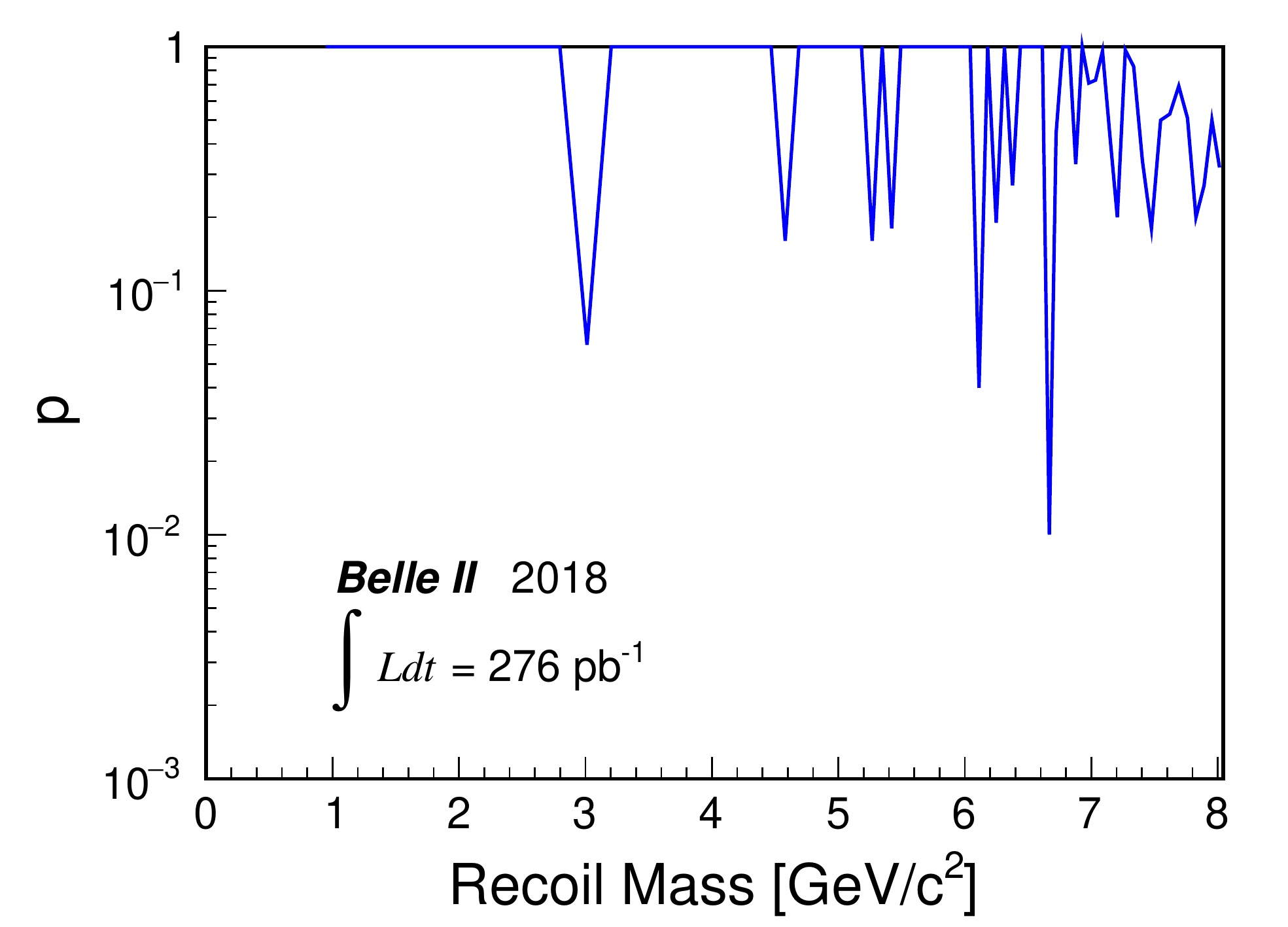}
    \caption{Probability $p$ to observe a number of events greater than the measured one given the predicted background level as a function of the recoil mass for the $\mu^+ \mu^-$ sample, evaluated with a half bin shift.}
    \label{fig:fig5}
\end{figure}

\begin{figure}[!ht]
    \centering
    \includegraphics[width=0.6\linewidth]{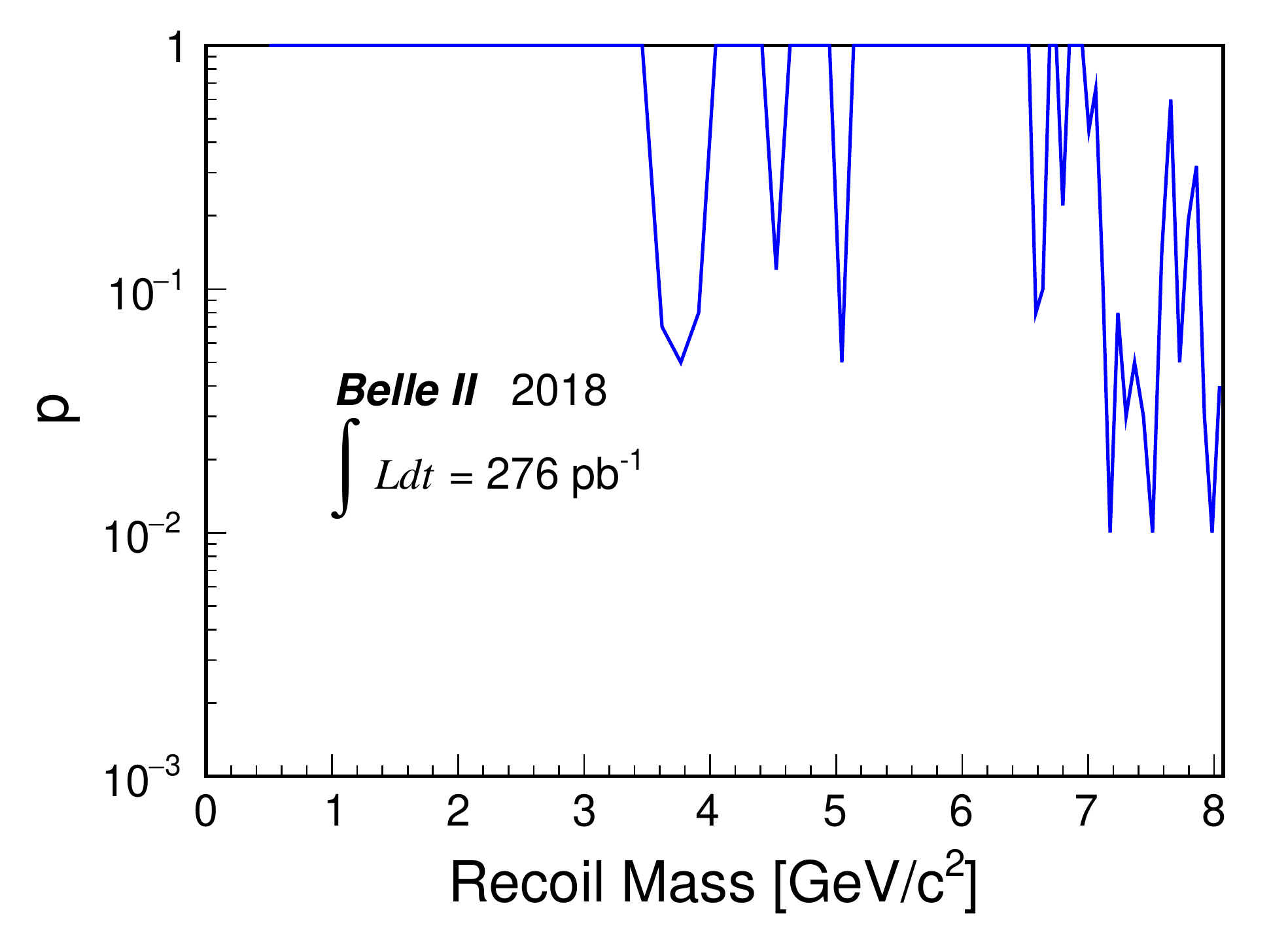}
    \caption{Probability $p$ to observe a number of events greater than the measured one given the predicted background level as a function of the recoil mass for the $e^{\pm} \mu^{\mp}$ sample.}
    \label{fig:fig6}
\end{figure}

\clearpage

\begin{longtable}{c|c|c|c|c|c}
  \caption{Standard $Z^{\prime}$ mass windows, observed yields $N_{\text{obs}}$, signal efficiency $\epsilon$, expected number of background events $N_{\text{bkg}}$, 90$\%$ CL upper limits on $\sigma(e^+ e^- \to \mu^+ \mu^- \text{invisible})$, and 90$\%$ CL upper limits on $g'$ (computed in the center of the $Z^{\prime}$ mass window) for BF$(Z^{\prime} \to \text{invisible})=1$.} \\
  
  \hline 
  $Z^{\prime}$ mass window [GeV/c$^2$] & $N_{\text{obs}}$ & $\epsilon$ & $N_{\text{bkg}}$ & $\sigma$ [fb] & $g'$ \\
  \hline
  \endfirsthead
  
  \multicolumn{6}{c}
  {\tablename\ \thetable\: continued from previous page} \\
  \hline
  $Z^{\prime}$ mass window [GeV/c$^2$] & $N_{\text{obs}}$ & $\epsilon$ & $N_{\text{bkg}}$ & $\sigma$ [fb] & $g^{\prime}$ \\
  \hline
  \endhead
  
  \hline
  \multicolumn{6}{r}{{\small Continued on next page}} \\
  \endfoot
  
  \hline
  \endlastfoot
  
    -0.150-1.150&0&0.028&0.438&316&0.051 \\
    1.150-1.632&0&0.049&0.213&178&0.070\\
    1.632-1.982&0&0.052&0.167&173&0.086\\
    1.982-2.287&0&0.052&0.145&169&0.100\\
    2.287-2.562&0&0.052&0.121&167&0.115\\
    2.562-2.805&0&0.052&0.092&167&0.129\\
    2.805-3.018&0&0.052&0.078&166&0.144\\
    3.018-3.208&1&0.051&0.073&290&0.209\\
    3.208-3.381&0&0.049&0.114&182&0.180\\
    3.381-3.542&0&0.047&0.104&188&0.198\\
    3.542-3.695&0&0.047&0.102&185&0.226\\
    3.695-3.840&0&0.048&0.136&187&0.254\\
    3.840-3.979&0&0.048&0.169&183&0.275\\
    3.979-4.110&0&0.048&0.151&181&0.296\\
    4.110-4.236&0&0.047&0.153&190&0.327\\
    4.236-4.355&0&0.046&0.160&197&0.356\\
    4.355-4.469&0&0.045&0.171&199&0.381\\
    4.469-4.580&0&0.044&0.182&202&0.408\\
    4.580-4.688&1&0.043&0.143&322&0.545\\
    4.688-4.794&0&0.043&0.186&204&0.461\\
    4.794-4.898&0&0.042&0.180&208&0.494\\
    4.898-4.998&0&0.042&0.206&207&0.520\\
    4.998-5.094&0&0.041&0.166&212&0.554\\
    5.094-5.184&0&0.040&0.264&218&0.590\\
    5.184-5.268&0&0.039&0.152&230&0.635\\
    5.268-5.347&1&0.039&0.175&366&0.833\\
    5.347-5.422&1&0.038&0.233&371&0.873\\
    5.422-5.493&0&0.037&0.200&232&0.723\\
    5.493-5.562&0&0.037&0.211&232&0.751\\
    5.562-5.630&0&0.036&0.190&247&0.803\\
    5.630-5.697&0&0.036&0.221&245&0.827\\
    5.697-5.765&0&0.035&0.291&254&0.874\\
    5.765-5.834&0&0.034&0.228&262&0.920\\
    5.834-5.903&0&0.034&0.252&265&0.960\\
    5.903-5.972&0&0.033&0.202&264&0.994\\
    5.972-6.042&0&0.032&0.196&272&$>1$\\
    6.042-6.112&1&0.031&0.374&434&$>1$\\
    6.112-6.180&1&0.030&0.194&474&$>1$\\
    6.180-6.248&1&0.030&0.268&472&$>1$\\
    6.248-6.313&0&0.029&0.237&308&$>1$\\
    6.313-6.377&1&0.028&0.342&487&$>1$\\
    6.377-6.439&0&0.027&0.248&329&$>1$\\
    6.439-6.499&0&0.026&0.225&330&$>1$\\
    6.499-6.557&0&0.026&0.186&327&$>1$\\
    6.557-6.613&0&0.027&0.354&320&$>1$\\
    6.613-6.668&2&0.028&0.367&690&$>1$\\
    6.668-6.721&2&0.028&0.537&662&$>1$\\
    6.721-6.773&0&0.029&0.860&314&$>1$\\
    6.773-6.825&0&0.030&0.766&287&$>1$\\
    6.825-6.876&1&0.030&0.959&413&$>1$\\
    6.876-6.927&1&0.031&1.176&388&$>1$\\
    6.927-6.979&0&0.032&1.203&278&$>1$\\
    6.979-7.032&3&0.033&2.004&582&$>1$\\
    7.032-7.087&2&0.038&6.069&318&$>1$\\
    7.087-7.144&12&0.044&11.357&778&$>1$\\
    7.144-7.204&16&0.050&15.229&804&$>1$\\
    7.204-7.267&22&0.057&21.406&885&$>1$\\
    7.267-7.334&12&0.064&26.799&425&$>1$\\
    7.334-7.403&36&0.071&32.765&1060&$>1$\\
    7.475-7.547&58&0.084&45.061&1413&$>1$\\
    7.547-7.620&40&0.082&47.553&836&$>1$\\
    7.620-7.691&44&0.081&49.545&959&$>1$\\
    7.691-7.761&42&0.079&49.448&883&$>1$\\
    7.761-7.827&62&0.077&51.369&1471&$>1$\\
    7.827-7.892&57&0.076&52.360&1345&$>1$\\
    7.892-7.953&58&0.074&50.480&1413&$>1$\\
    7.953-8.014&53&0.073&52.640&1287&$>1$\\
    8.014-8.072&64&0.071&54.018&1566&$>1$

\end{longtable}

\clearpage

\begin{longtable}{c|c|c|c}
  \caption{LFV $Z^{\prime}$ mass window, observed yields $N_{\text{obs}}$, expected number of background events $N_{\text{bkg}}$, 90$\%$ CL upper limits on efficiency times cross section $\epsilon \times \sigma(e^+ e^- \to e^{\pm} \mu^{\mp} \text{invisible})$.} \\
  
  \hline 
  $Z^{\prime}$ mass window [GeV/c$^2$] & $N_{\text{obs}}$ & $N_{\text{bkg}}$ & $\epsilon \times \sigma$ [fb] \\
  \hline
  \endfirsthead
  
  \multicolumn{4}{c}
  {\tablename\ \thetable\: continued from previous page} \\
  \hline
  $Z^{\prime}$ mass window [GeV/c$^2$] & $N_{\text{obs}}$ & $N_{\text{bkg}}$ & $\epsilon \times \sigma$ [fb] \\
  \hline
  \endhead
  
  \hline
  \multicolumn{4}{r}{{\small Continued on next page}} \\
  \endfoot
  
  \hline
  \endlastfoot
  
    -0.150-1.150 & 0 & 0.000-0.010 & 9.62 \\
    1.150-1.632 & 0 & 0.018 & 9.67\\
    1.632-1.982 & 0 & 0.030 & 9.64\\
    1.982-2.287 & 0 & 0.036 & 9.64\\
    2.287-2.562 & 0 & 0.006 & 9.62\\
    2.562-2.805 & 0 & 0.048 & 9.62\\
    2.805-3.018 & 0 & 0.030 & 9.60\\
    3.018-3.208 & 0 & 0.065 & 9.62\\
    3.208-3.381 & 0 & 0.068 & 9.66\\
    3.381-3.542 & 0 & 0.153 & 9.65\\
    3.542-3.695 & 1 & 0.096 & 16.36\\
    3.695-3.840 & 1 & 0.078 & 16.41\\
    3.840-3.979 & 1 & 0.117 & 16.41\\
    3.979-4.110 & 0 & 0.163 & 9.61\\
    4.110-4.236 & 0 & 0.077 & 9.63\\
    4.236-4.355 & 0 & 0.145 & 9.65\\
    4.355-4.469 & 0 & 0.115 & 9.63\\
    4.469-4.580 & 1 & 0.172 & 16.20\\
    4.580-4.688 & 0 & 0.083 & 9.62\\
    4.688-4.794 & 0 & 0.139 & 9.63\\
    4.794-4.898 & 0 & 0.107 & 9.66\\
    4.898-4.998 & 0 & 0.131 & 9.64\\
    4.998-5.094 & 1 & 0.065 & 16.46\\
    5.094-5.184 & 0 & 0.151 & 9.64\\
    5.184-5.268 & 0 & 0.143 & 9.65\\
    5.268-5.347 & 0 & 0.178 & 9.64\\
    5.347-5.422 & 0 & 0.131 & 9.64\\
    5.422-5.493 & 0 & 0.172 & 9.65\\
    5.493-5.562 & 0 & 0.143 & 9.63\\
    5.562-5.630 & 0 & 0.187 & 9.63\\
    5.630-5.697 & 0 & 0.167 & 9.62\\
    5.697-5.765 & 0 & 0.177 & 9.61\\
    5.765-5.834 & 0 & 0.101 & 9.62\\
    5.834-5.903 & 0 & 0.151 & 9.65\\
    5.903-5.972 & 0 & 0.243 & 9.61\\
    5.972-6.042 & 0 & 0.127 & 9.61\\
    6.042-6.112 & 0 & 0.133 & 9.63\\
    6.112-6.180 & 0 & 0.113 & 9.64\\
    6.180-6.248 & 0 & 0.151 & 9.64\\
    6.248-6.313 & 0 & 0.098 & 9.62\\
    6.313-6.377 & 0 & 0.149 & 9.61\\
    6.377-6.439 & 0 & 0.053 & 9.64\\
    6.439-6.499 & 0 & 0.080 & 9.63\\
    6.499-6.557 & 0 & 0.098 & 9.62\\
    6.557-6.613 & 1 & 0.109 & 16.32\\
    6.613-6.668 & 1 & 0.151 & 16.20\\
    6.668-6.721 & 0 & 0.172 & 9.64\\
    6.721-6.773 & 0 & 0.207 & 9.64\\
    6.773-6.825 & 1 & 0.338 & 15.66\\
    6.825-6.876 & 0 & 0.310 & 9.65\\
    6.876-6.927 & 0 & 0.461 & 9.64\\
    6.927-6.979 & 0 & 0.450 & 9.65\\
    6.979-7.032 & 1 & 0.849 & 14.30\\
    7.032-7.087 & 3 & 4.905 & 16.66\\
    7.087-7.144 & 10 & 8.327 & 32.04\\
    7.144-7.204 & 18 & 11.817 & 54.09\\
    7.204-7.267 & 19 & 16.685 & 46.06\\
    7.267-7.334 & 24 & 19.384 & 57.71\\
    7.334-7.403 & 26 & 22.504 & 58.24\\
    7.403-7.475 & 31 & 25.539 & 69.66\\
    7.475-7.547 & 34 & 26.179 & 77.82\\
    7.547-7.620 & 26 & 26.301 & 52.73\\
    7.620-7.691 & 16 & 23.941 & 31.35\\
    7.691-7.761 & 26 & 22.620 & 59.28\\
    7.761-7.827 & 19 & 19.767 & 41.71\\
    7.827-7.892 & 15 & 17.611 & 34.88\\
    7.892-7.953 & 20 & 15.609 & 51.65\\
    7.953-8.014 & 20 & 13.736 & 56.78\\
    8.014-8.072 & 16 & 12.557 & 44.62

\end{longtable}

%% file: zprime_authors.tex

\newcommand{\instSinica}{Academia Sinica, Taipei 11529}
\newcommand{\instBeihang}{Beihang University, Beijing 100191}
\newcommand{\instBUAP}{Benemerita Universidad Autonoma de Puebla, Puebla 72570}
\newcommand{\instBNL}{Brookhaven National Laboratory, Upton, New York 11973}
\newcommand{\instBINP}{Budker Institute of Nuclear Physics SB RAS, Novosibirsk 630090}
\newcommand{\instCMU}{Carnegie Mellon University, Pittsburgh, Pennsylvania 15213}
\newcommand{\instCinvestavIPN}{Centro de Investigacion y de Estudios Avanzados del Instituto Politecnico Nacional, Mexico City 07360}
\newcommand{\instPrague}{Faculty of Mathematics and Physics, Charles University, 121 16 Prague}
\newcommand{\instChiangMai}{Chiang Mai University, Chiang Mai 50202}
\newcommand{\instChiba}{Chiba University, Chiba 263-8522}
\newcommand{\instChonnam}{Chonnam National University, Gwangju 61186}
\newcommand{\instConacyt}{Consejo Nacional de Ciencia y Tecnolog\'{\i}a, Mexico City 03940}
\newcommand{\instDESY}{Deutsches Elektronen--Synchrotron, 22607 Hamburg}
\newcommand{\instDuke}{Duke University, Durham, North Carolina 27708}
\newcommand{\instDuyTan}{Institute of Theoretical and Applied Research (ITAR), Duy Tan University, Hanoi 100000, Vietnam}
\newcommand{\instENEA}{ENEA Casaccia, I-00123 Roma}
\newcommand{\instEri}{Earthquake Research Institute, University of Tokyo, Tokyo 113-0032}
\newcommand{\instJuelich}{Forschungszentrum J\"{u}lich, 52425 J\"{u}lich}
\newcommand{\instFuJen}{Department of Physics, Fu Jen Catholic University, Taipei 24205}
\newcommand{\instFudan}{Key Laboratory of Nuclear Physics and Ion-beam Application (MOE) and Institute of Modern Physics, Fudan University, Shanghai 200443}
\newcommand{\instGoettingen}{II. Physikalisches Institut, Georg-August-Universit\"{a}t G\"{o}ttingen, 37073 G\"{o}ttingen}
\newcommand{\instGifu}{Gifu University, Gifu 501-1193}
\newcommand{\instSOKENDAI}{The Graduate University for Advanced Studies (SOKENDAI), Hayama 240-0193}
\newcommand{\instGyeongsang}{Gyeongsang National University, Jinju 52828}
\newcommand{\instHanyang}{Department of Physics and Institute of Natural Sciences, Hanyang University, Seoul 04763}
\newcommand{\instKEK}{High Energy Accelerator Research Organization (KEK), Tsukuba 305-0801}
\newcommand{\instJPARC}{J-PARC Branch, KEK Theory Center, High Energy Accelerator Research Organization (KEK), Tsukuba 305-0801}
\newcommand{\instIISER}{Indian Institute of Science Education and Research Mohali, SAS Nagar, 140306}
\newcommand{\instIITBhubaneswar}{Indian Institute of Technology Bhubaneswar, Satya Nagar 751007}
\newcommand{\instIITGuwahati}{Indian Institute of Technology Guwahati, Assam 781039}
\newcommand{\instIITHyderabad}{Indian Institute of Technology Hyderabad, Telangana 502285}
\newcommand{\instIITMadras}{Indian Institute of Technology Madras, Chennai 600036}
\newcommand{\instIndiana}{Indiana University, Bloomington, Indiana 47408}
\newcommand{\instIHEPRussia}{Institute for High Energy Physics, Protvino 142281}
\newcommand{\instHEPHYVienna}{Institute of High Energy Physics, Vienna 1050, Austria}
\newcommand{\instIHEPChina}{Institute of High Energy Physics, Chinese Academy of Sciences, Beijing 100049}
\newcommand{\instChennai}{Institute of Mathematical Sciences, Chennai 600113}
\newcommand{\instIPP}{Institute of Particle Physics (Canada), Victoria, British Columbia V8W 2Y2}
\newcommand{\instIOP}{Institute of Physics, Hanoi}
\newcommand{\instIFIC}{Instituto de Fisica Corpuscular, Paterna 46980}
\newcommand{\instFrascati}{INFN Laboratori Nazionali di Frascati, I-00044 Frascati}
\newcommand{\instNapoliINFN}{INFN Sezione di Napoli, I-80126 Napoli}
\newcommand{\instPadovaINFN}{INFN Sezione di Padova, I-35131 Padova}
\newcommand{\instPerugiaINFN}{INFN Sezione di Perugia, I-06123 Perugia}
\newcommand{\instPisaINFN}{INFN Sezione di Pisa, I-56127 Pisa}
\newcommand{\instRomaINFN}{INFN Sezione di Roma, I-00185 Roma}
\newcommand{\instRomaTreINFN}{INFN Sezione di Roma Tre, I-00146 Roma}
\newcommand{\instTorinoINFN}{INFN Sezione di Torino, I-10125 Torino}
\newcommand{\instTriesteINFN}{INFN Sezione di Trieste, I-34127 Trieste}
\newcommand{\instJAEA}{Advanced Science Research Center, Japan Atomic Energy Agency, Naka 319-1195}
\newcommand{\instMainz}{Johannes Gutenberg-Universit\"{a}t Mainz, Institut f\"{u}r Kernphysik, D-55099 Mainz}
\newcommand{\instGiessen}{Justus-Liebig-Universit\"{a}t Gie\ss{}en, 35392 Gie\ss{}en}
\newcommand{\instKarlsruhe}{Institut f\"{u}r Experimentelle Teilchenphysik, Karlsruher Institut f\"{u}r Technologie, 76131 Karlsruhe}
\newcommand{\instKennesaw}{Kennesaw State University, Kennesaw, Georgia 30144}
\newcommand{\instKitasato}{Kitasato University, Sagamihara 252-0373}
\newcommand{\instKISTI}{Korea Institute of Science and Technology Information, Daejeon 34141}
\newcommand{\instKorea}{Korea University, Seoul 02841}
\newcommand{\instKSU}{Kyoto Sangyo University, Kyoto 603-8555}
\newcommand{\instKyotoU}{Kyoto University, Kyoto 606-8501}
\newcommand{\instKyungpook}{Kyungpook National University, Daegu 41566}
\newcommand{\instLAL}{Laboratoire de l'Acc\'{e}l\'{e}rateur Lin\'{e}aire, IN2P3/CNRS et Universit\'{e} Paris-Sud 11, Centre Scientifique d'Orsay, F-91898 Orsay Cedex}
\newcommand{\instLPI}{P.N. Lebedev Physical Institute of the Russian Academy of Sciences, Moscow 119991}
\newcommand{\instLNNU}{Liaoning Normal University, Dalian 116029}
\newcommand{\instLMU}{Ludwig Maximilians University, 80539 Munich}
\newcommand{\instLuther}{Luther College, Decorah, Iowa 52101}
\newcommand{\instMNITJaipur}{Malaviya National Institute of Technology Jaipur, Jaipur 302017}
\newcommand{\instMPP}{Max-Planck-Institut f\"{u}r Physik, 80805 M\"{u}nchen}
\newcommand{\instMPGHLL}{Semiconductor Laboratory of the Max Planck Society, 81739 M\"{u}nchen}
\newcommand{\instMcGill}{McGill University, Montr\'{e}al, Qu\'{e}bec, H3A 2T8}
\newcommand{\instMETU}{Middle East Technical University, 06531 Ankara}
\newcommand{\instMIPT}{Moscow Institute of Physics and Technology, Moscow Region 141700}
\newcommand{\instMEPhI}{Moscow Physical Engineering Institute, Moscow 115409}
\newcommand{\instNagoya}{Graduate School of Science, Nagoya University, Nagoya 464-8602}
\newcommand{\instNagoyaKMI}{Kobayashi-Maskawa Institute, Nagoya University, Nagoya 464-8602}
\newcommand{\instNaraWu}{Nara Women's University, Nara 630-8506}
\newcommand{\instUNAM}{National Autonomous University of Mexico, Mexico City}
\newcommand{\instNTUTaiwan}{Department of Physics, National Taiwan University, Taipei 10617}
\newcommand{\instNUUTaiwan}{National United University, Miao Li 36003}
\newcommand{\instKrakow}{H. Niewodniczanski Institute of Nuclear Physics, Krakow 31-342}
\newcommand{\instNiigata}{Niigata University, Niigata 950-2181}
\newcommand{\instNSU}{Novosibirsk State University, Novosibirsk 630090}
\newcommand{\instOkinawa}{Okinawa Institute of Science and Technology, Okinawa 904-0495}
\newcommand{\instOsakaCity}{Osaka City University, Osaka 558-8585}
\newcommand{\instRCNP}{Research Center for Nuclear Physics, Osaka University, Osaka 567-0047}
\newcommand{\instPNNL}{Pacific Northwest National Laboratory, Richland, Washington 99352}
\newcommand{\instPanjab}{Panjab University, Chandigarh 160014}
\newcommand{\instPeking}{Peking University, Beijing 100871}
\newcommand{\instPanjabPAU}{Punjab Agricultural University, Ludhiana 141004}
\newcommand{\instRIKEN}{Theoretical Research Division, Nishina Center, RIKEN, Saitama 351-0198}
\newcommand{\instXavier}{St. Francis Xavier University, Antigonish, Nova Scotia, B2G 2W5}
\newcommand{\instSeoul}{Seoul National University, Seoul 08826}
\newcommand{\instShandong}{Shandong University, Jinan 250100}
\newcommand{\instSPU}{Showa Pharmaceutical University, Tokyo 194-8543}
\newcommand{\instSoochow}{Soochow University, Suzhou 215006}
\newcommand{\instSoongsil}{Soongsil University, Seoul 06978}
\newcommand{\instLjubljanaJSI}{J. Stefan Institute, 1000 Ljubljana}
\newcommand{\instKyiv}{Taras Shevchenko National Univ. of Kiev, Kiev}
\newcommand{\instTata}{Tata Institute of Fundamental Research, Mumbai 400005}
\newcommand{\instTUM}{Department of Physics, Technische Universit\"{a}t M\"{u}nchen, 85748 Garching}
\newcommand{\instECUTUM}{Excellence Cluster Universe, Technische Universit\"{a}t M\"{u}nchen, 85748 Garching}
\newcommand{\instTelAviv}{Tel Aviv University, School of Physics and Astronomy, Tel Aviv, 69978}
\newcommand{\instToho}{Toho University, Funabashi 274-8510}
\newcommand{\instTohoku}{Department of Physics, Tohoku University, Sendai 980-8578}
\newcommand{\instTitech}{Tokyo Institute of Technology, Tokyo 152-8550}
\newcommand{\instTokyoMetropolitan}{Tokyo Metropolitan University, Tokyo 192-0397}
\newcommand{\instUAS}{Universidad Autonoma de Sinaloa, Sinaloa 80000}
\newcommand{\instNapoliUNIV}{Dipartimento di Scienze Fisiche, Universit\`{a} di Napoli Federico II, I-80126 Napoli}
\newcommand{\instNapoliUNIVA}{Dipartimento di Agraria, Universit\`{a} di Napoli Federico II, I-80055 Portici (NA)}
\newcommand{\instPadovaUNIV}{Dipartimento di Fisica e Astronomia, Universit\`{a} di Padova, I-35131 Padova}
\newcommand{\instPerugiaUNIV}{Dipartimento di Fisica, Universit\`{a} di Perugia, I-06123 Perugia}
\newcommand{\instPisaUNIV}{Dipartimento di Fisica, Universit\`{a} di Pisa, I-56127 Pisa}
\newcommand{\instRomaUNIV}{Universit\`{a} di Roma ``La Sapienza,'' I-00185 Roma}
\newcommand{\instRomaTreUNIV}{Dipartimento di Matematica e Fisica, Universit\`{a} di Roma Tre, I-00146 Roma}
\newcommand{\instTorinoUNIV}{Dipartimento di Fisica, Universit\`{a} di Torino, I-10125 Torino}
\newcommand{\instTriesteUNIV}{Dipartimento di Fisica, Universit\`{a} di Trieste, I-34127 Trieste}
\newcommand{\instMontreal}{Universit\'{e} de Montr\'{e}al, Physique des Particules, Montr\'{e}al, Qu\'{e}bec, H3C 3J7}
\newcommand{\instIPHC}{Universit\'{e} de Strasbourg, CNRS, IPHC, UMR 7178, 67037 Strasbourg}
\newcommand{\instAdelaide}{Department of Physics, University of Adelaide, Adelaide, South Australia 5005}
\newcommand{\instBonn}{University of Bonn, 53115 Bonn}
\newcommand{\instUBC}{University of British Columbia, Vancouver, British Columbia, V6T 1Z1}
\newcommand{\instCincinnati}{University of Cincinnati, Cincinnati, Ohio 45221}
\newcommand{\instFlorida}{University of Florida, Gainesville, Florida 32611}
\newcommand{\instHamburg}{University of Hamburg, 20148 Hamburg}
\newcommand{\instHawaii}{University of Hawaii, Honolulu, Hawaii 96822}
\newcommand{\instHeidelberg}{University of Heidelberg, 68131 Mannheim}
\newcommand{\instLjubljanaUniLJ}{Faculty of Mathematics and Physics, University of Ljubljana, 1000 Ljubljana}
\newcommand{\instLouisville}{University of Louisville, Louisville, Kentucky 40292}
\newcommand{\instMalaya}{National Centre for Particle Physics, University Malaya, 50603 Kuala Lumpur}
\newcommand{\instLjubljanaUM}{University of Maribor, 2000 Maribor}
\newcommand{\instMelbourne}{School of Physics, University of Melbourne, Victoria 3010}
\newcommand{\instMississippi}{University of Mississippi, University, Mississippi 38677}
\newcommand{\instUOM}{University of Miyazaki, Miyazaki 889-2192}
\newcommand{\instNovaGorica}{University of Nova Gorica, 5000 Nova Gorica}
\newcommand{\instPittsburgh}{University of Pittsburgh, Pittsburgh, Pennsylvania 15260}
\newcommand{\instUSTC}{University of Science and Technology of China, Hefei 230026}
\newcommand{\instSAlabama}{University of South Alabama, Mobile, Alabama 36688}
\newcommand{\instSCarolina}{University of South Carolina, Columbia, South Carolina 29208}
\newcommand{\instSydney}{School of Physics, University of Sydney, New South Wales 2006}
\newcommand{\instTabuk}{Department of Physics, Faculty of Science, University of Tabuk, Tabuk 71451}
\newcommand{\instUTokyo}{Department of Physics, University of Tokyo, Tokyo 113-0033}
\newcommand{\instIPMU}{Kavli Institute for the Physics and Mathematics of the Universe (WPI), University of Tokyo, Kashiwa 277-8583}
\newcommand{\instVictoria}{University of Victoria, Victoria, British Columbia, V8W 3P6}
\newcommand{\instVPI}{Virginia Polytechnic Institute and State University, Blacksburg, Virginia 24061}
\newcommand{\instWayneState}{Wayne State University, Detroit, Michigan 48202}
\newcommand{\instYamagata}{Yamagata University, Yamagata 990-8560}
\newcommand{\instYerevan}{Alikhanyan National Science Laboratory, Yerevan 0036}
\newcommand{\instYonsei}{Yonsei University, Seoul 03722}

\affiliation{\instBeihang}
\affiliation{\instBNL}
\affiliation{\instBINP}
\affiliation{\instCMU}
\affiliation{\instCinvestavIPN}
\affiliation{\instPrague}
\affiliation{\instChiba}
\affiliation{\instChonnam}
\affiliation{\instConacyt}
\affiliation{\instDESY}
\affiliation{\instDuke}
\affiliation{\instDuyTan}
\affiliation{\instEri}
\affiliation{\instJuelich}
\affiliation{\instFuJen}
\affiliation{\instFudan}
\affiliation{\instGifu}
\affiliation{\instSOKENDAI}
\affiliation{\instGyeongsang}
\affiliation{\instHanyang}
\affiliation{\instKEK}
\affiliation{\instJPARC}
\affiliation{\instIITHyderabad}
\affiliation{\instIITMadras}
\affiliation{\instIndiana}
\affiliation{\instIHEPRussia}
\affiliation{\instHEPHYVienna}
\affiliation{\instIHEPChina}
\affiliation{\instIPP}
\affiliation{\instIOP}
\affiliation{\instIFIC}
\affiliation{\instFrascati}
\affiliation{\instNapoliINFN}
\affiliation{\instPadovaINFN}
\affiliation{\instPerugiaINFN}
\affiliation{\instPisaINFN}
\affiliation{\instRomaINFN}
\affiliation{\instRomaTreINFN}
\affiliation{\instTorinoINFN}
\affiliation{\instTriesteINFN}
\affiliation{\instJAEA}
\affiliation{\instMainz}
\affiliation{\instGiessen}
\affiliation{\instKarlsruhe}
\affiliation{\instKitasato}
\affiliation{\instKISTI}
\affiliation{\instKorea}
\affiliation{\instKyungpook}
\affiliation{\instLAL}
\affiliation{\instLPI}
\affiliation{\instLMU}
\affiliation{\instLuther}
\affiliation{\instMNITJaipur}
\affiliation{\instMPP}
\affiliation{\instMcGill}
\affiliation{\instMIPT}
\affiliation{\instMEPhI}
\affiliation{\instNagoya}
\affiliation{\instNagoyaKMI}
\affiliation{\instNaraWu}
\affiliation{\instNTUTaiwan}
\affiliation{\instNUUTaiwan}
\affiliation{\instKrakow}
\affiliation{\instNiigata}
\affiliation{\instNSU}
\affiliation{\instOkinawa}
\affiliation{\instOsakaCity}
\affiliation{\instRCNP}
\affiliation{\instPNNL}
\affiliation{\instPanjab}
\affiliation{\instPeking}
\affiliation{\instPanjabPAU}
\affiliation{\instRIKEN}
\affiliation{\instSeoul}
\affiliation{\instSPU}
\affiliation{\instSoongsil}
\affiliation{\instLjubljanaJSI}
\affiliation{\instKyiv}
\affiliation{\instTata}
\affiliation{\instTUM}
\affiliation{\instTelAviv}
\affiliation{\instTohoku}
\affiliation{\instTitech}
\affiliation{\instTokyoMetropolitan}
\affiliation{\instUAS}
\affiliation{\instNapoliUNIV}
\affiliation{\instPadovaUNIV}
\affiliation{\instPerugiaUNIV}
\affiliation{\instPisaUNIV}
\affiliation{\instRomaTreUNIV}
\affiliation{\instTorinoUNIV}
\affiliation{\instTriesteUNIV}
\affiliation{\instMontreal}
\affiliation{\instIPHC}
\affiliation{\instBonn}
\affiliation{\instUBC}
\affiliation{\instCincinnati}
\affiliation{\instHawaii}
\affiliation{\instLjubljanaUniLJ}
\affiliation{\instLouisville}
\affiliation{\instMalaya}
\affiliation{\instLjubljanaUM}
\affiliation{\instMelbourne}
\affiliation{\instMississippi}
\affiliation{\instUOM}
\affiliation{\instPittsburgh}
\affiliation{\instUSTC}
\affiliation{\instSAlabama}
\affiliation{\instSCarolina}
\affiliation{\instSydney}
\affiliation{\instUTokyo}
\affiliation{\instIPMU}
\affiliation{\instVictoria}
\affiliation{\instVPI}
\affiliation{\instWayneState}
\affiliation{\instYamagata}
\affiliation{\instYerevan}
\affiliation{\instYonsei}

  \author{I.~Adachi}\affiliation{\instKEK}\affiliation{\instSOKENDAI} 
  \author{P.~Ahlburg}\affiliation{\instBonn} 
  \author{H.~Aihara}\affiliation{\instUTokyo} 
  \author{N.~Akopov}\affiliation{\instYerevan} 
  \author{A.~Aloisio}\affiliation{\instNapoliUNIV}\affiliation{\instNapoliINFN} 
  \author{N.~Anh~Ky}\affiliation{\instIOP}\affiliation{\instDuyTan} 
  \author{D.~M.~Asner}\affiliation{\instBNL} 
  \author{H.~Atmacan}\affiliation{\instCincinnati} 
  \author{T.~Aushev}\affiliation{\instMIPT} 
  \author{V.~Aushev}\affiliation{\instKyiv} 
% ...

\collaboration{Belle II Collaboration}

%% file: zprime_acknowledgements.tex
We thank the SuperKEKB group for the excellent operation of the
accelerator; the KEK cryogenics group for the efficient
operation of the solenoid; and the KEK computer group for
on-site computing support.
This work was supported by the following funding sources:
Science Committee of the Republic of Armenia Grant No.~18T-1C180;
Australian Research Council and research grant Nos.
DP180102629, 
DP170102389, 
DP170102204, 
DP150103061, 
FT130100303, 
and
FT130100018; 
Austrian Federal Ministry of Education, Science and Research, and
Austrian Science Fund No.~P 31361-N36; 
Natural Sciences and Engineering Research Council of Canada, Compute Canada and CANARIE;
Chinese Academy of Sciences and research grant No.~QYZDJ-SSW-SLH011,
National Natural Science Foundation of China and research grant Nos.
11521505,
11575017,
11675166,
11761141009,
11705209,
and
11975076,
LiaoNing Revitalization Talents Program under contract No.~XLYC1807135,
Shanghai Municipal Science and Technology Committee under contract No.~19ZR1403000,
Shanghai Pujiang Program under Grant No.~18PJ1401000,
and the CAS Center for Excellence in Particle Physics (CCEPP);
the Ministry of Education, Youth and Sports of the Czech Republic under Contract No.~LTT17020 and 
Charles University grants SVV 260448 and GAUK 404316;
European Research Council, 7th Framework PIEF-GA-2013-622527, 
Horizon 2020 Marie Sklodowska-Curie grant agreement No.~700525 `NIOBE,' 
Horizon 2020 Marie Sklodowska-Curie RISE project JENNIFER grant agreement No.~644294,
Horizon 2020 ERC-Advanced Grant No.~267104, and
NewAve No.~638528 (European grants); 
L'Institut National de Physique Nucl\'{e}aire et de Physique des Particules (IN2P3) du CNRS (France);
BMBF, DFG, HGF, MPG and AvH Foundation (Germany);
Department of Atomic Energy and Department of Science and Technology (India);
Israel Science Foundation grant No.~2476/17
and
United States-Israel Binational Science Foundation grant No.~2016113;
Istituto Nazionale di Fisica Nucleare and the research grants BELLE2;
Japan Society for the Promotion of Science,  Grant-in-Aid for Scientific Research grant Nos.
16H03968, 
16H03993, 
16H06492,
16K05323, 
17H01133, 
17H05405, 
18K03621, 
18H03710, 
18H05226,
19H00682, 
26220706,
and
26400255,
the National Institute of Informatics, and Science Information NETwork 5 (SINET5), 
and
the Ministry of Education, Culture, Sports, Science, and Technology (MEXT) of Japan;  
National Research Foundation (NRF) of Korea Grant Nos.
2016R1\-D1A1B\-01010135,
2016R1\-D1A1B\-02012900,
2018R1\-A2B\-3003643,
2018R1\-A4A\-1025334,
2018R1\-A6A1A\-06024970,
2018R1\-D1A1B\-07047294,
2019K1\-A3A7A\-09033840,
and
2019R1\-I1A3A\-01058933,
Radiation Science Research Institute,
Foreign Large-size Research Facility Application Supporting project,
the Global Science Experimental Data Hub Center of the Korea Institute of Science and Technology Information
and
KREONET/GLORIAD;
Universiti Malaya RU grant, Akademi Sains Malaysia and Ministry of Education Malaysia;
Frontiers of Science Program contracts
FOINS-296,
CB-221329,
CB-236394,
CB-254409,
and
CB-180023, and the Thematic Networks program (Mexico);
the Polish Ministry of Science and Higher Education and the National Science Center;
the Ministry of Science and Higher Education of the Russian Federation,
Agreement 14.W03.31.0026;
Slovenian Research Agency and research grant Nos.
J1-9124
and
P1-0135; 
Agencia Estatal de Investigacion, Spain grant Nos.
FPA2014-55613-P
and
FPA2017-84445-P,
and
CIDEGENT/2018/020 of Generalitat Valenciana;
Ministry of Science and Technology and research grant Nos.
MOST106-2112-M-002-005-MY3
and
MOST107-2119-M-002-035-MY3, 
and the Ministry of Education (Taiwan);
Thailand Center of Excellence in Physics;
TUBITAK ULAKBIM (Turkey);
Ministry of Education and Science of Ukraine;
the US National Science Foundation and research grant Nos.
PHY-1807007 
and
PHY-1913789, 
and the US Department of Energy and research grant Nos.
DE-AC06-76RLO1830, 
DE-SC0007983, 
DE-SC0009824, 
DE-SC0009973, 
DE-SC0010073, 
DE-SC0010118, 
DE-SC0010504, 
DE-SC0011784, 
DE-SC0012704; 
and
the National Foundation for Science and Technology Development (NAFOSTED) 
of Vietnam under contract No 103.99-2018.45.